\documentclass{article}

\usepackage[english]{babel}

\usepackage[a4paper]{geometry}

\usepackage{amsmath}
\usepackage{physics}

\usepackage{graphicx}
\usepackage[colorlinks=true, allcolors=blue]{hyperref}

\usepackage[super,sort&compress,comma]{natbib}
 
\usepackage{caption}
\usepackage{subcaption}

\usepackage{authblk}

\title{A precision method for integrating shock sensors in the lining of sports helmets by additive manufacturing}

\author[1]{Aferdita Xhameni}
\author[2]{Runbei Cheng}
\author[3,*]{Tristan Farrow}
\affil[1]{Department of Electronic \& Electrical Engineering, Malet Place, University College London, WC1E 7JE, United Kingdom}
\affil[2]{Department of Engineering, University of Oxford, Parks Road, Oxford, OX1 3PJ, United Kingdom}
\affil[3]{Clarendon Laboratory, Department of Physics, University of Oxford, Parks Road, Oxford OX1 3PU, United Kingdom}
\affil[*]{Corresponding author: Tristan Farrow (email: tristan.farrow@physics.ox.ac.uk)}

\date{}

\begin{document}
\maketitle

%\begin{abstract}
\textbf{
\noindent
A method is presented for embedding sensors into the lining of sports helmets for the purpose of monitoring head impact in contact sports. Additive manufacturing was used to embed optical fibre-based pressure sensors inside thermoplastic polyurethane (TPU), a material used for lining and shock-dampening that fills the space between a helmet’s hard outer shell and the head. A proof-of-concept has successfully demonstrated sensitive shock-monitoring capabilities, thus avoiding the inaccuracies in existing systems. The sensors can be embedded into additively manufactured parts in new or existing helmets and are unobtrusive to the wearers. This proposed system can be powered by readily-available medical-grade cell batteries and requires less maintenance.
}\leavevmode\newline
%\end{abstract}

%need linebreak here without using "newline" or "\\"


{A}{thletes} with a history of repetitive head impacts and multiple traumatic brain injuries (TBI) risk long-term neurological sequels and premature cognitive decline. Protective headgear is a crucial factor in reducing head trauma, but there are still a significant number of injuries every year with 1.7-3 million concussions recorded annually across sports in the US \cite{covassin2018sex}, with 300,000 in all sports \cite{smith2015ice} and approximately one third of these occuring during American football games \cite{centers1997sports}\cite{thurman1998epidemiology}. Research has shown that regular small impacts to the skull can lead to long-term brain damage, with a 10-20\% prevalence in retired NFL players \cite{casson2014there}. There is a need for objective monitoring methods that can better help to safeguard athletes.
Head impact monitoring can be valuable for providing data to develop safer techniques during training which can help players avoid collisions, as well as support medical professionals determine when to withdraw a player during a competition. Existing systems involve accelerometers and gyroscopes but have yet to demonstrate their viability in the trial phase because they can suffer from inaccuracy. Low-intensity and detrimental impact can go undetected, while the force of a collision can often be overestimated by 6\% \cite{beckwith2012measuring}\cite{patton2016review}. Additionally, existing helmet-sensor solutions require replacing batteries in a factory at least once over the course of an American football season, and mouthguard systems can suffer from poor coupling to the head during impact when athletes' jaws relax \cite{patton2016review}.

Fibre Bragg Grating sensors (FBGs) are versatile optical sensors that can measure various physical properties, such as strain and temperature. They are widely used in fields such as aerospace and civil engineering due to their robustness to environmental factors such as moisture, rust, and electromagnetic interference, unlike their electrical counterparts. FBGs can achieve sampling rates in the range of 100kHz \cite{bentell2009250}, which is excellent compared to electrical sensors, which sample at about 10kHz \cite{patton2016review}. Unlike electrical sensors, FBGs can measure strain along a single direction, making them more suitable for structural health monitoring and healthcare applications \cite{tosi2018review}. In low power, portable strain sensing applications, FBGs have been used to measure strain using LED light sources powered by medical grade batteries at less than 0.25W \cite{maheshwari2019chirped}.  

FBGs consist of sections with different refractive indices called Bragg reflectors. The Bragg reflectors can reflect light at a specific wavelength, depending on the refractive index, as well as on the size and separation of the grating. When the fibre changes length due to either an applied tension or heat expansion, the grating separation changes, causing the reflected (Bragg) wavelength to shift. The total shift due to strain and temperature can be modelled as \cite{betz2006advanced}

 \begin{equation}
 \label{noise_model}
     \frac{\Delta\lambda_B}{\lambda_{B0}} = [1-P^{eff}]\epsilon_z^{s,m} + [(1-P^{eff})\alpha^s + \frac{1}{n_0}\frac{dn}{dT}]\Delta T,
 \end{equation}

  where $\Delta\lambda_B$ is the change in Bragg wavelength from the initial Bragg wavelength $\lambda_{B0}$, $[1-P^{eff}]$ is the strain sensitivity of the FBG, $\epsilon_z^{s,m}$ is the mechanical strain due to tension, $\alpha^s \Delta T$ is the thermal strain due to thermal expansion, and $n_0$ is the initial refractive index.

When attached to mechanical objects, FBGs can be used to detect deformation and vibration. The accuracy of such measurements depends on the bonding strength between the FBGs and the measured objects. In practice, FBGs are often glued to the surface of the objects of interest. However, embedding FBGs inside objects can create stronger bonds and allow the objects’ structural health to be monitored more accurately and with more information \cite{maier2012embedded}.

%\IEEEpubidadjcol
Additive Manufacturing, commonly known as 3D printing, offers the freedom to rapidly prototype complex structures. Most additive manufacturing printing methods produce parts by laying down materials layer by layer. This mechanism allows the possibility of embedding FBGs inside printed structures during the printing process. A well-established method for embedding FBGs into additively manufactured metal structures has already been realized \cite{schomer2016characterization}. However, there have only been few attempts to embed them into 3D printed polymer structures with varying degrees of success \cite{maier2012embedded}\cite{zubel2016embedding}\cite{manzo2019embedding}. 
    
Two of the most common polymer additive manufacturing methods are fused filament fabrication (FFF) and stereolithography apparatus (SLA). 3D printers employing these methods are widely available commercially and affordable. Most modern machines can achieve layer resolutions in the sub-millimetre range, with more advanced models being capable of 50µm and 25µm layer resolution for FFF and SLA, respectively. With such a resolution, it is possible to produce complex parts with securely embedded FBGs reproducibly. Being able to manufacture such objects with relative ease through additive manufacturing creates new possibilities in a wide range of fields. Hence, it is of great interest to standardize and validate such embedding processes.

While optical fibres are tough and resilient to harsh conditions, the engraved sections of FBGs are, however, very fragile in comparison and can be damaged when high heat is applied during the embedding process \cite{maier2012embedded}\cite{schomer2016characterization}. This temperature restriction eliminates most metal additive manufacturing methods as candidates for FBG embedding. One low temperature 3D printing method is Ultrasonic Additive Manufacturing (UAM), where metal sheets are welded together using high power ultrasound \cite{schomer2016characterization}. UAM has been adapted as the industry gold standard method for embedding FBGs in metal parts used for high-value tasks by organizations like NASA but remains prohibitively expensive for niche consumer products \cite{hehr2018integrating}. Integrating FBG sensors in helmets and other consumer applications via polymer additive manufacturing (FFF, SLA) as shown here presents an opportunity for cost-effective manufacturing at scale.

\section*{Embedding Methods}
\subsection*{Fused Filament Fabrication}
%Text here
FFF is a layer-by-layer printing method using thermoplastic materials, polymers that can be reshaped with heat. During printing, thermoplastic filaments are pushed through a heated nozzle, which softens the filaments upon contact, turning the filaments into a malleable liquid. This liquid is then deposited onto a build platform layer by layer and quickly solidified by a cooling fan, forming predesigned parts. There are a large variety of materials available for FFF printing, such as polylactic acid (PLA), acrylonitrile butadiene styrene (ABS), high-impact polystyrene (HIPS), thermoplastic polyurethane (TPU), and nylon, making it ideal for printing functional parts.

During FFF prints, polymer filaments leave the print nozzle in a softened liquid-like state. Thus, FBGs can be embedded by placing the fibre on an unfinished surface halfway through the print. The softened polymer adheres to the fibre forming a tight bond between the fibre and the printed parts. FFF print nozzles generally operate at ~200°C, and such high heat could damage the fibre if the nozzle were to make direct contact with the fibre \cite{maier2012embedded}\cite{schomer2016characterization}\cite{grobnic2021fiber}\cite{yan2017fabrication}. This is commonly mitigated by designing the printed parts with grooves for fibre placement (fig. 1). In addition to protecting the fibre from the print nozzle, this also facilitates a precise alignment of the fibre.

\begin{figure}[h!]
    \centering
    \includegraphics[scale=0.38]{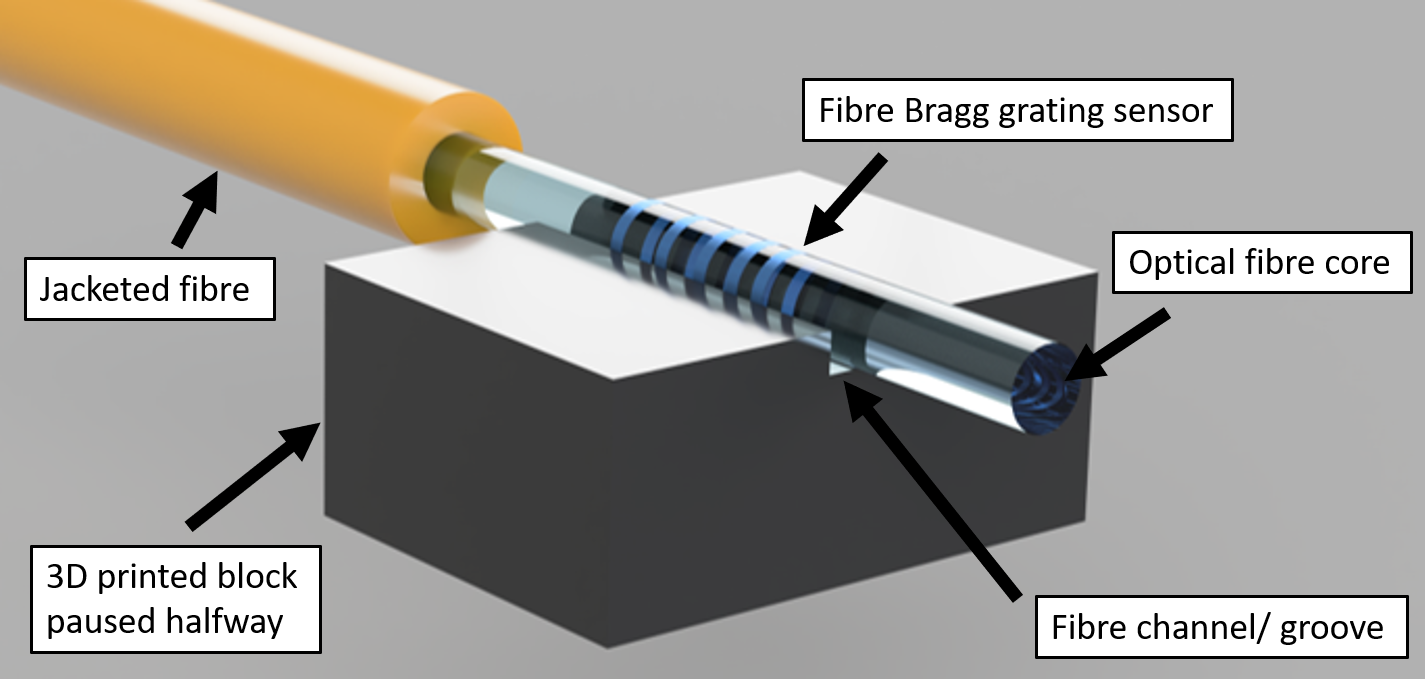}
    \caption{Embedding FBGs in square-faced grooves was a limitation of the printer layer resolution, however the melted polymer provides good adhesion to the fibre. This technique is commonly used in additive manufacturing \cite{zubel2016embedding}\cite{manzo2019embedding}.}
    \label{fig:my_label}
\end{figure}

\subsection*{Stereolithography}

Unlike FFF which reshapes solid plastic filaments to produce parts, SLA uses photopolymer resin to produce higher-quality prints. To produce prints with this method, an upside-down build platform is submerged into a resin tank. The bottom of the resin tank has a UV light source, with wavelength \textasciitilde400nm, that cures the resin onto the build platform, layer by layer. Prints are only partially cured fresh out of the printer, and a further UV bath is required to cure them fully. SLA machines have an even more extensive selection of materials than FFF, ranging from hard materials such as ceramic to flexible materials like rubber, offering unique solutions to a wide range of fields.

The challenges with using these printers to embed FBGs are the spatial limitations imposed by the resin tank and screen wiper, as well as the upside-down geometry of the build platform. We modified the platform such that, during a print, the fibre is held in place against the drag from the viscous resin and aligned in the groove, without damage.

This method performed consistently for multiple test prints with silica fibres. The fibres were securely embedded inside the SLA blocks without fail with the correct groove size. However, when polymer (polyimide) fibres were used, the resins used could no longer cure fully, even after extensive exposure to UV in the post curing UV bath. Polyimide has a low acid resistance while SLA resins are acidic. Their unfavourable chemical interaction is likely to have released contaminants into the resin and inhibited curing. More research is needed to investigate this behaviour in-depth to control it.

In addition to the interaction between the resin and polyimide, there are other concerns with this method. Photopolymer resin is highly toxic. Partially cured and uncured resin can poison the operator if mishandled. Additionally, photopolymer resin is cured using a UV light source, and since FBG fibres are also photosensitive to the UV spectrum, FBG fibres could potentially be damaged during the printing and post curing processes. A preliminary UV damage test was carried out during optical testing, the results of which are discussed below.

%\section*{\label{sec:setup}Results}
\section*{Results}
%Text here
To evaluate the embedded sensors, $10 \time 10 \time 50 mm^{3}$ cuboid prototypes with optical fibres embedded were manufactured and tested. Two types of optical fibres were embedded, plain fibres and FBGs. The plain fibres were 242µm in diameter, and the FBGs were 125µm in core diameter at the grating, and 242µm in diameter throughout the rest of the fibre. 125µm is the smallest layer thickness compatible with the FFF machine, and 50µm is the smallest layer thickness for the SLA machine. Both were used to maximize the print quality.
\begin{figure}[h!]
    \centering
    \includegraphics[scale=0.38]{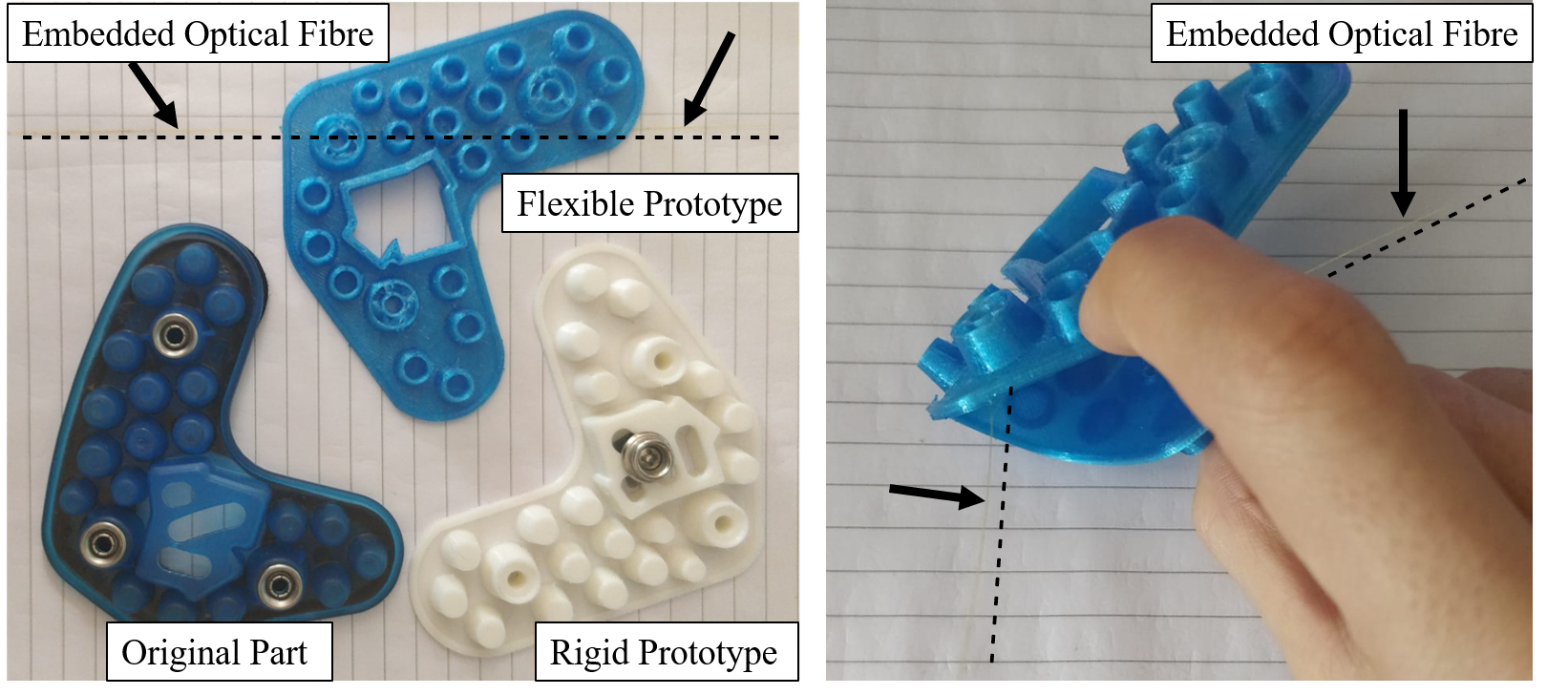}
    \caption{Optical fibre embedded in flexible TPU and rigid PLA parts modelled after helmet padding segment. The fibre can bend and flex with the material without damage or slipping.}
    \label{fig:my_label}
\end{figure}

Optical fibres were also embedded in a rigid and flexible cheek pad (fig. 2) modelled from a Schutt Vengeance Pro American football helmet (Schutt Sports, Litchfield, Illinois, United States) using an FFF printer. Such segments can be printed for the whole helmet, comprising a sensor array.   

\subsection*{Optical Analysis}
%Text here
A LUNA 4600 optical backscatter reflectometer (OBR) (Luna Innovations, Roanoke, Virginia, United States) was used to measure signal amplitude as a function of the distance travelled by the signal in the fibre. Comparing a 2cm long FBG embedded by FFF to the reference signal from a 5cm long non-embedded FBG (fig. 3), we observe a square wave signal in both cases. %The embedded FBG is 3cm shorter than the reference FBG, hence the width of the signal is also shorter. 
%This is a result of the manufacturing method causing tension to be held in the fibre due to the pinch points at the ends of the block (fig. 6b). 

\begin{figure}[h!]
    \centering
    \includegraphics[scale=0.35]{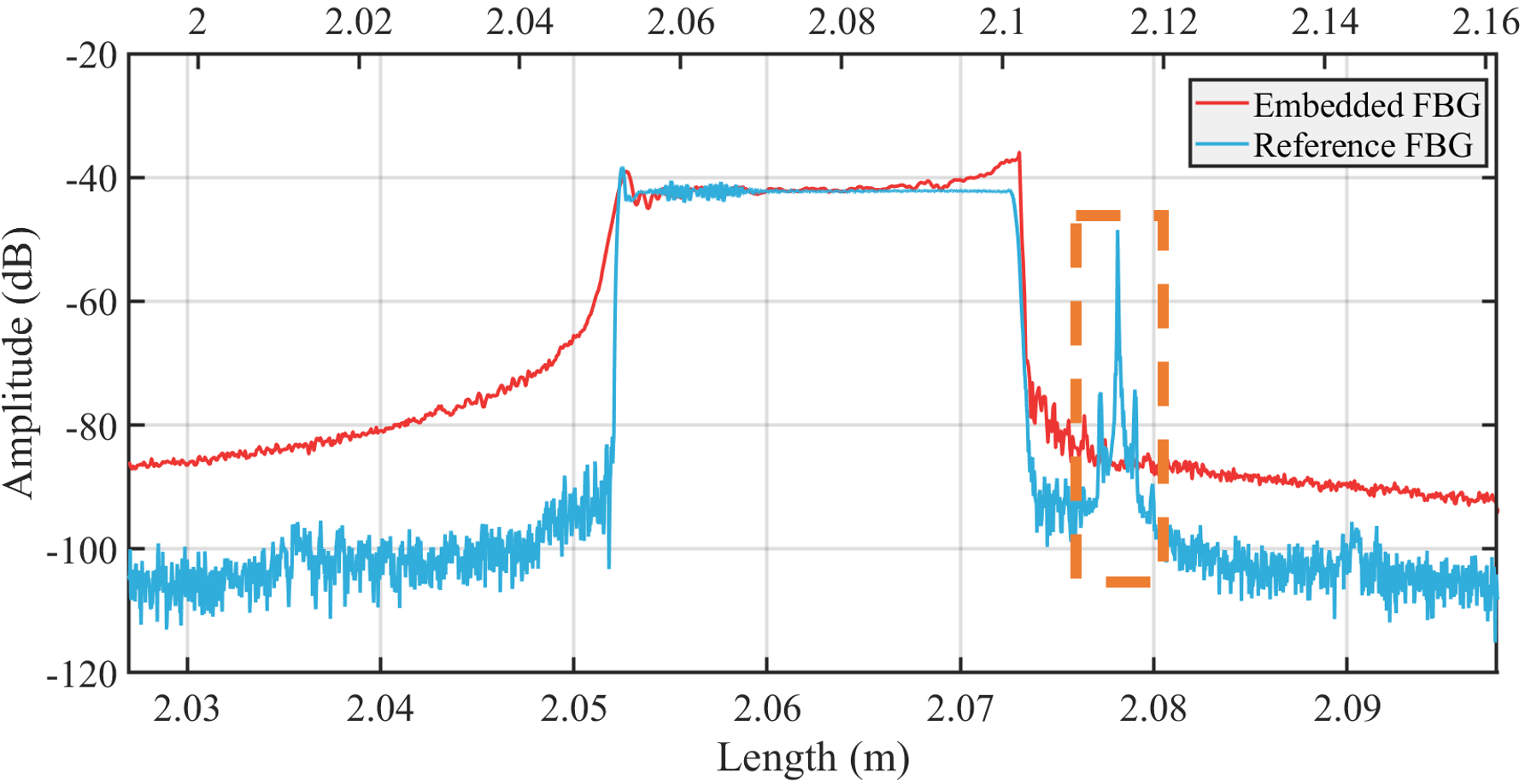}
    \caption{The embedded FBG (bottom axis) is 3cm shorter than the non - embedded reference FBG (top axis). The absence of an additional peak highlighted in orange for the embedded FBG resulted from removing the reflective end cap in the FBG for ease of handling.}
    \label{fig:my_label}
\end{figure}

High temperatures in the range of ~200°C can completely extinguish the change in refractive index in the bare FBG \cite{yan2017fabrication}. However, the high amplitude reflections in the FBG generating the square-wave signal are still present after embedding. Hence, we can conclude that temperature did not damage the grating during embedding.  
 
In figure 3, the embedded signal amplitude (red) rises gradually compared to the sharp rise in the reference signal (blue). We hypothesise that the tension is held in the embedded portion of the fibre due to the pinch points (fig. 6b) located at each end of the printed block. The strain this induces in the fibre changes the period of the grating and results in weak reflections of wavelengths outside the design range. These show up as flatter tails in the tails of the square wave signals from embedded fibres (fig. 4).

\begin{figure}[h!]
    \centering
    \includegraphics[scale=0.35]{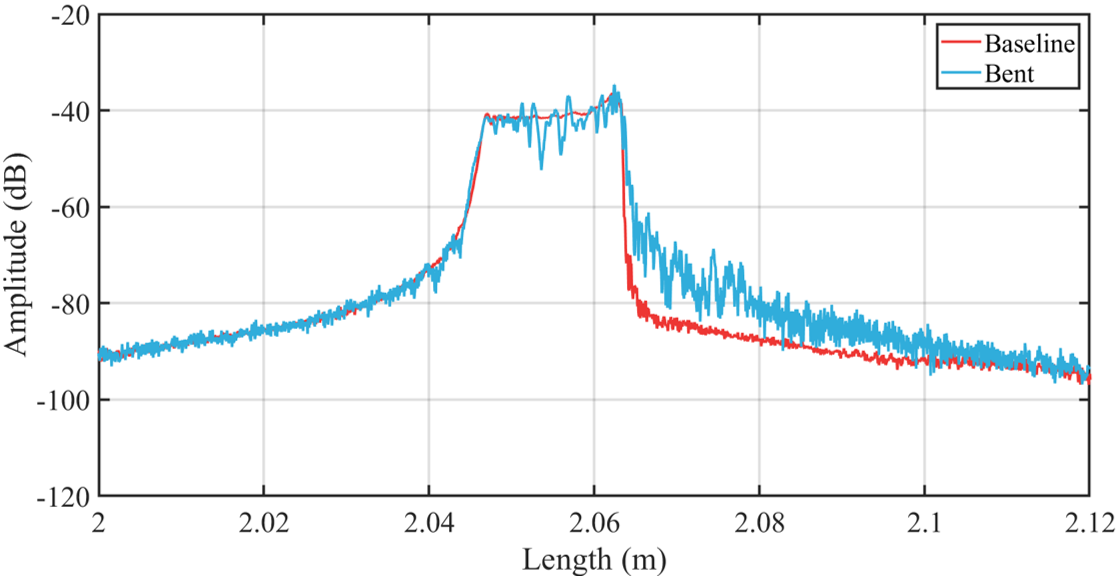}
    \caption{Strain induced in the FBG as a result of bending the FBG results in a similar deformation of the signal as embedding the FBG with pinch points (fig. 6b).}
    \label{fig:my_label}
\end{figure}

To confirm this, we coiled FBGs around cylinders and recorded the reflected signal after bending, as this would result in a similar extension of the grating under tension. We observe that tension under bending increases the curvature in the signal, compared to the signal prior to bending. This led us to confirm our hypothesis that tension is held in the embedded fibre at pinch points where it enters and exits the block. The issue can be minimized by printing support blocks on both ends of the structure, at the same height as the groove.

We exposed a silica coated FBG embedded by FFF to the same UV bath used in the SLA process to investigate the effects of 405nm UV on the gratings (fig. 5). %Although this would provide insight,  this is not an exact representation of all the conditions a fibre is exposed to during the embedding process.

\begin{figure}[h!]
    \centering
    \includegraphics[scale=0.35]{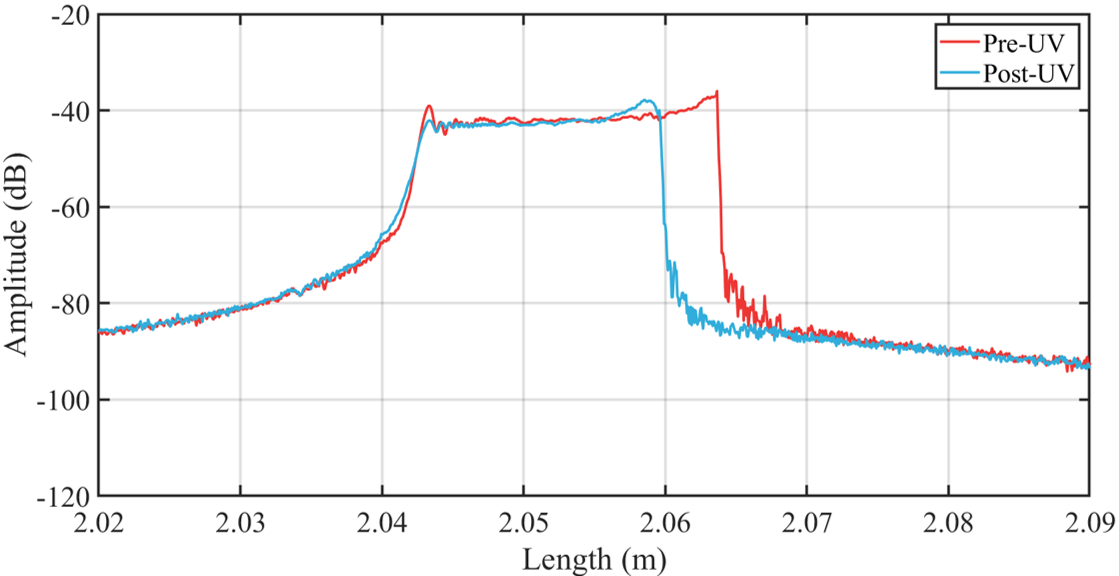}
    \caption{To gauge the effect of 405nm UV light used in SLA printing on the sensor, an FBG was left in a UV bath of the same wavelength. A portion of the sensor was snapped off due to mishandling, resulting in the length variation seen in this figure. The characteristic curve of the sensor has not changed otherwise. Completely erasing a portion of the curve by UV exposure can only be achieved by illuminating each grating element in a chirped FBG, at a displacement of one half-period \cite{leng2017uv}. It is unlikely that unfocused UV light in a water bath would be capable of this. Thus, it is likely that 405nm UV light used in SLA printing is not harmful to FBG sensors.%Caption about broken region and UV not doing any damage.
    }
    \label{fig:my_label}
\end{figure}
 As the whole fibre is exposed to unfocused UV light from all angles during curing, any UV damage would degrade the characteristic FBG signal entirely \cite{leng2017uv}. This was not observed hence it is unlikely that any localised UV damage occurred which would affect only a portion of the signal.

\subsection*{Mechanical Testing}
%\textcolor{orange}{Rather than phrasing this section as estimating the upper bound of the sensor's measuring capability, we should rephrase it to testing the adhesion between the sensors and the printed parts. This also means we need to change parts of the discussion respectively. And we should add references for how civil engineers have used embedded FBGs to monitor strain/presure, and say that we can now do the same with helmets using our embedding method in the discussion.} 
To measure the strain experienced by the structure the FBG is embedded in, there needs to be a high degree of adhesion at the interface between them. %To test the adhesion strength of FBGs embedded in additively manufactured parts,  
A hanging-weights experiment was carried out to measure the workload required to cause slippage between the embedded fibres and the printed parts (fig. 6a). FFF printed PLA and TPU blocks embedded with silica-coated fibres were used. The blocks were clamped to a metal pole with the fibre hanging vertically under the block while masses were incrementally added. A section of the fibre was allowed to protrude the top of the block and was marked as a reference point to measure the slippage. The experiment showed that the assembly was slippage-free up to 1 kg for PLA and TPU. %Since the fibre did not slip from the block up to this point, 
Thus, it is likely that the sensor and surrounding material experience the same strain.

\begin{figure}[h!]
    \centering
        \includegraphics[scale=0.18]{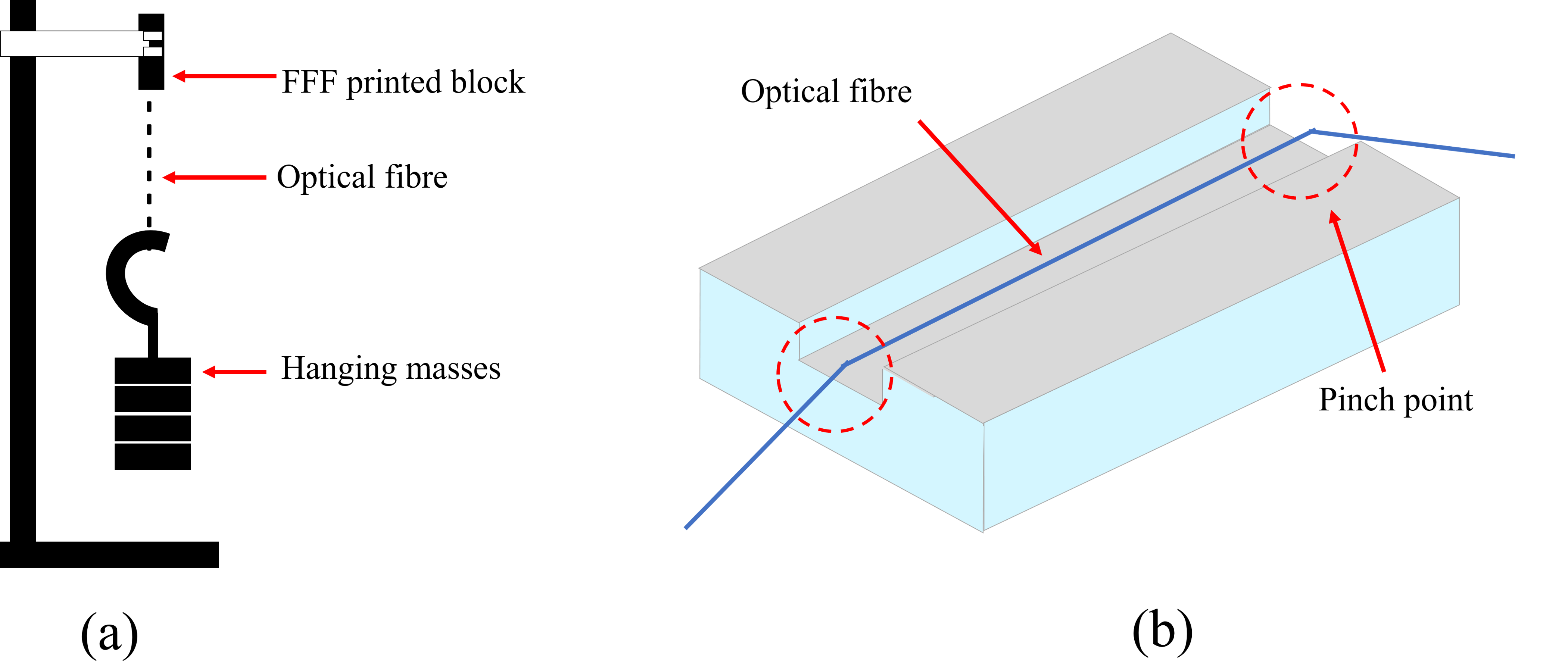}
        \caption{(a) A hanging mass experiment was conducted to test the adhesion of the fibre to the printed part. (b) Pinch points where the fibre exits the printed part can hold tension in the embedded fibre.}
    \label{fig:my_label}
\end{figure}

 %Using the Young’s modulus of the fibre and the block, an upper bound for measurable workload on the block was calculated as follows. FFF printed PLA has a Young’s modulus around 2.5 GPa \cite{zhao2019novel}, and a coated silica optical fibre has a Young’s modulus around 17 GPa \cite{antunes2012mechanical}; the printed block has a 10×10 mm2 cross section, and the coated fibre has a diameter of 0.25 mm. Using the Young’s modulus equation: Young's modulus = (Applied Force)/(Strain × Cross-section Area)
%Given that the block and the fibre experienced the same strain, the upper bound for a measurable workload on the printed block is estimated to ~30 MPa.

\section*{Discussion}

\label{sec:furtherwork}

%update figure, change names and package better

FBGs are versatile sensors that are reliable and resilient under demanding environmental conditions. By embedding them inside complex structures, they can be used to monitor vibration and structural health \cite{liu2007temperature}\cite{manzo2019embedding}. This study validated the use of FFF to embed FBGs inside complex polymer structures such as padding elements for American football helmets. %By including a groove for fibre placement in the design, fibres can easily be aligned in a specific direction and are protected from the hot printer nozzle. %Our prototype can measure strain accurately up to a pressure of $\sim 30$ MPa applied to the printed structure, and is cheap to manufacture at scale.

In addition to validating the potential for embedding FBGs with FFF, we demonstrated a new technique using SLA. While SLA delivers higher quality prints than FFF, it cannot be used to embed polyimide fibres due to their chemical interaction with the SLA resin. SLA is more costly than FFF and has a longer print time, which are trade-offs that need to be balanced against the higher quality prints depending on the application use-case.

In summary, this study paves the way for field-testing prototypes of FBGs embedded in the lining of sports helmets under in-situ workloads and real-time strain response. Prototypes with multiple fibres for temperature correction can also be fabricated to enhance accuracy where needed. Other types of sensors can be embedded using 3D printing, such as coaxial cables, for example, using SLA, which we embedded during our development phase. With printable FBGs on the horizon \cite{hong2019design}, we anticipate the printing of polymer structures with built-in FBG sensors for a range of new applications that need precise in-situ monitoring, such as the structural health of materials used in aviation, civil and medical engineering in life-saving technologies.

\section*{Competing interests} TF, AX, and RC have filed GB patent applications no. 2107898.5 and PCT/GB2022/051410.

\section*{Acknowledgment}
TF acknowledges support from the Gordon and Betty Moore Foundation.

\bibliography{references.bib}
\bibliographystyle{unsrt}

\end{document}